\documentclass[twocolumn,english,amsart,showpacs,preprintnumbers,amsmath,amssymb,floatfix]{revtex4-1}
\usepackage{tikz,xcolor}
\usepackage[colorlinks = true,
linkcolor = blue,
urlcolor  = blue,
citecolor = blue,
anchorcolor = blue]{hyperref}
\definecolor{lime}{HTML}{A6CE39}
\DeclareRobustCommand{\orcidicon}{%
	\begin{tikzpicture}
		\draw[lime, fill=lime] (0,0)
		circle [radius=0.16]
		node[white] {{\fontfamily{qag}\selectfont \tiny ID}};
		\draw[white, fill=white] (-0.0625,0.095)
		circle [radius=0.007];
	\end{tikzpicture}
	\hspace{-2mm}
}
\foreach \x in {A, ..., Z}{%
	\expandafter\xdef\csname orcid\x\endcsname{\noexpand\href{https://orcid.org/\csname orcidauthor\x\endcsname}{\noexpand\orcidicon}}
}

\usepackage[T1]{fontenc}
\usepackage[latin9]{inputenc}
\usepackage{color}
\usepackage{array}
\usepackage{amstext}
\usepackage{graphicx}
\usepackage{esint}
\usepackage{rotating}
\usepackage[font=small,labelfont=bf]{caption}
\usepackage{xcolor}

\makeatletter

\@ifundefined{textcolor}{}
{%
 \definecolor{BLACK}{gray}{0}
 \definecolor{WHITE}{gray}{1}
 \definecolor{RED}{rgb}{1,0,0}
 \definecolor{GREEN}{rgb}{0,1,0}
 \definecolor{BLUE}{rgb}{0,0,1}
 \definecolor{CYAN}{cmyk}{1,0,0,0}
 \definecolor{MAGENTA}{cmyk}{0,1,0,0}
 \definecolor{YELLOW}{cmyk}{0,0,1,0}
 }

\@ifundefined{definecolor}
 {\usepackage{color}}{}
\@ifundefined{definecolor}
 {\usepackage{color}}{}
\makeatother
\usepackage{babel}

\begin{document}

%%%%%%%%%%%%%%%%%%%%%%%%%%%%%%

\title { QCD analysis of $xF_3$ structure functions in deep-inelastic scattering: Mellin transform by Gegenbauer polynomial up to N$^3$LO approximation  }

\author {Fatemeh Arbabifar$^{1}$\orcidA{}}
\email{F.Arbabifar@cfu.ac.ir}

\author{Nader Morshedian$^{2}$\orcidC{}}
\email{nmorshed@aeoi.org.ir}

\author{Leila Ghasemzadeh$^{3}$\orcidD{}}
\email{leila.ghasemzadeh71@gmail.com}

\author {Shahin Atashbar Tehrani$^{4,5}$\orcidB{}}
\email{Atashbar@ipm.ir}

\affiliation {
$^{(1)}$Department of Physics Education, Farhangian University, P.O.Box 14665-889, Tehran, Iran.   \\
$^{(2)}$Plasma and Nuclear Fusion Research School, Nuclear Science and Technology Research Institute, P.O.Box 14399-51113, Tehran, Iran.\\
$^{(3)}$Physics Department, Yazd University, P.O.Box 89195-741, Yazd, Iran\\
$^{(4)}$School of Particles and Accelerators, Institute for Research in Fundamental Sciences (IPM), P.O.Box 19395-5531, Tehran, Iran.\\
$^{(5)}$Department of Physics, Faculty of Nano and Bio Science and Technology, Persian Gulf University, 75169 Bushehr, Iran. 
}

\date{\today}

%
%%%%%%%%%%%%%%%%%%%%%%%%%%%%%%%%%%%%%%%%%%%%%%%%%%%%%%%%%%%%%%%%%%%%%%%%%%%%%%%%%%%%%%%%%%%%%%%%%%%%%%%
\begin{abstract}\label{abstract}

{ 
 This paper provides a thorough examination of the $xF_3$ structure functions in deep-inelastic scattering through a comprehensive QCD analysis. Our approach harnesses sophisticated mathematical techniques, namely the Mellin transform combined with Gegenbauer polynomials.  We have employed the Jacobi polynomials approach for analysis, conducting investigations at three levels of precision: Next-to-Leading Order (NLO), Next-to-Next-to-Leading Order (N$^2$LO), and Next-Next-Next-to-Leading Order (N$^3$LO). We have performed a comparison of our sets of valence-quark parton distribution functions with those of recent research groups, specifically CT18 and MSHT20 at NLO and N$^2$LO, and MSTH23 at N$^3$LO, which are concurrent with our current analysis. The combination of Mellin transforms with Gegenbauer polynomials proves to be a powerful tool for investigating the $xF_3$ structure functions in deep-inelastic scattering and the results obtained from our analysis demonstrate a favorable alignment with experimental data.

}

\end{abstract}
%

%\pacs{12.39.-x, 14.65.Bt, 12.38.-t, 12.38.Bx}
\maketitle
%\tableofcontents{}

%
%%%%%%%%%%%%%%%%%%%%%%%%%%%%%%%%%%%%%%%%%%%%%%%%%%%%%%%%%%%%%%%%%%%%%%%%%%%%%%%%%%%%%%%%%%%%%%%
%
\section{Introduction}\label{sec:intro}
Quantum Chromodynamics (QCD) is the fundamental theory of strong interactions, describing the behavior of quarks and gluons, the building blocks of protons, neutrons, and other hadrons. Deep-Inelastic Scattering (DIS) experiments have been a cornerstone in the study of QCD, providing crucial insights into the internal structure of nucleons and the distribution of quarks and gluons within them. Among the various observables in DIS, the $xF_3$ structure functions hold particular significance as they encode essential information about the parton distribution functions (PDFs) within the nucleon.

The $xF_3$ structure functions, related to the  charged current DIS, play a pivotal role in testing the predictions of QCD and probing the dynamics of quarks and gluons at different energy scales. The study of $xF_3$~\cite{MoosaviNejad:2016ebo,Khorramian:2006wg,Tooran:2019cfz,Sidorov:2015bea,Sidorov:2014aca,Nath:2011sv,Kataev:2002rj,Kataev:2002wr,Kataev:2001kk,Santiago:2001mh,Alekhin:1999af,Kataev:1999bp,Sidorov:1997oia,Kataev:1997vv,Kataev:1997nc,Sidorov:1996if,Kataev:1998ce,Kataev:1996vu,Kataev:1994rj,Martin:1986gv,Behring:2015roa} has witnessed significant progress over the years, with advances in both experimental techniques and theoretical frameworks. To extract precise information from experimental data and interpret it in the context of QCD, sophisticated theoretical tools and mathematical techniques are required.

In this context, the Mellin transform~\cite{Gluck:1989ze}, which allows for the determination of moments of the proton structure functions, has emerged as an effective tool for the quantum chromodynamics (QCD) analysis of the $xF_3$ structure functions. The Mellin transform facilitates the improvement of approximations, leading to more accurate predictions for scaling violations and the evolution of parton distribution functions (PDFs) with changing momentum scales.  We have used Jacobi polynomials to perform the transformation of the evolved functions from the Mellin space to the Bjorken $x$ space, which is an important step in our analysis, as it allows us to compare our results with experimental data. Additionally, we leverage Gegenbauer polynomials for PDFs parameterization, which offer orthogonality and flexibility in function approximation. We show that using the Gegenbauer polynomial expansion method and the next-to-next-to-next-to-leading order (NNNLO) approximation, which are novel and accurate techniques in this context. 

Using Gegenbauer polynomials in models can increase the precision and quality of the approximations and the theoretical models. The best values for the parameters of these polynomials can be chosen by fitting them to experimental data. 

This paper presents a comprehensive QCD analysis of the proton structure function $xF_3$ using neutrino-nucleus scattering data from CCFR~\cite{Seligman:1997mc}, NuTeV~\cite{NuTeV:2005wsg} and CHORUS~\cite{CHORUS:2005cpn} experiments. We explain the theoretical framework of QCD and the Jacobi polynomial approach, and the key concepts for the subsequent analysis.

We conclude that the QCD analysis of the $xF_3$ structure function using Mellin transforms with Jacobi polynomials and Gegenbauer polynomials for PDFs parameterization.This methods is a fast, precise, and direct way to calculate the final structure function with high accuracy.  It allows us to extract the valence-quark distribution functions from the neutrino-nucleus scattering data without complicated
calculations in the kinematic region of interest.

This article is organized as follows. In Sec.~\ref{sec:sec2}, we introduce the PDFs parametrization and the theoretical framework for the Mellin transform. In Sec.~\ref{sec:sec4}, we describe the Jacobi polynomials and the QCD fit procedure. In Sec.~\ref{sec:sec6}, we present the fit results. In Sec.~\ref{sec:sec7}, we discuss the GLS sum rule. The final section, Sec.~\ref{sec:sec8}, summarizes and concludes the article.

\section{ Theoretical framework to Mellin transform }\label{sec:sec2}

In charged-current neutrino deep inelastic scattering (DIS) processes, a neutrino $\nu$ ($\bar{\nu}$) interacts with a quark inside the nucleon through the exchange of a virtual $W^\pm$ boson. The nonsinglet structure function $xF_3(x, Q^2)$, which arises from the parity-violating weak interaction, characterizes the momentum density of partons, including both valence quarks and antiquarks, within the nucleon.

	In the quark-parton model (QPM), the structure functions  $xF_3^{\nu p}$  and  $xF_3^{\bar \nu p}$  for neutrino-proton and antineutrino-proton interactions are given by changing the signs of the antiquark distributions in the expressions for  $F_2^{\nu p}$  and  $F_2^{\bar \nu p}$ . By considering $F_2   = 2xF_1$,  one can have the above $F_2^{\nu p}$ and  $F_2^{\bar \nu p}$ structure functions in terms of PDFs, $F_2^{\nu p}   = 2x(d + s + \bar u + \bar c)$,  and $F_2^{\bar \nu p} =2x(u + c + \bar d + \bar s)$.  By changing the signs of $\bar u$, $\bar d$, $\bar s$, and $\bar c$, The correct expressions for  $xF_3^{\nu p}$  and  $xF_3^{\bar \nu p}$  are:
\begin{eqnarray}
	xF_3^{\nu p} & =& 2x(d + s - \bar u - \bar c)~,   \nonumber   \\ 
	xF_3^{\bar \nu p} & =& 2x(u + c - \bar d - \bar s)~.   
\end{eqnarray}
By considering $u\equiv u_v+\bar u$ and $d\equiv d_v+\bar d$ and combining the above equations,  the structure function $xF_3$ is as follows:
\begin{equation}
	xF_3^{{({\nu}+{\bar \nu})}p}=xF_3^{\nu p}  + xF_3^{\bar \nu p}  
	= 2x(u_v  + d_v ) + 2x(s - \bar s) + 2x(c - \bar c)~.
\end{equation}
So, one can have the average of the neutrino and antineutrino nucleon structure function as follows:
\begin{eqnarray}
	xF_3^{N}(x,Q^2)&=&\frac{1}{2}\left(xF_3^{\nu N}+ xF_3^{\bar \nu N} \right) (x,Q^2)\nonumber \\
		&=&\frac{1}{2}\left([xF_3^{(\nu+\bar \nu) p}+xF_3^{(\nu+\bar \nu) n}]/2 \right)(x,Q^2)~.\nonumber\\
\end{eqnarray}
However, due to the isospin symmetry, $xF_3^{{({\nu}+{\bar \nu})}p}=xF_3^{{({\nu}+{\bar \nu})}n}$,  the  average of the neutrino and antineutrino nucleon structure is  
\begin{eqnarray}
	xF_3^{N}(x,Q^2)&=&\frac{1}{2}~xF_3^{(\nu+\bar \nu) p}(x,Q^2)\nonumber \\
	&&= [x(u_v  + d_v ) + x(s - \bar s) + x(c - \bar c)](x,Q^2)~.\nonumber\\
\end{eqnarray}

It is important to recognize that the differences between the strange quark and its antiquark $s-\bar{s}$, as well as the charm quark and its antiquark $c-\bar{c}$, are typically negligible. Consequently, the average structure of the nucleon as probed by neutrinos and antineutrinos predominantly reflects the distribution of valence quarks
\begin{eqnarray}
	xF_3^{N}(x,Q^2)&= (xu_v + xd_v) (x,Q^2)~.
\end{eqnarray}

where the combinations $d_v \equiv d - \bar{d}$ and $u_v \equiv u - \bar{u}$ correspond to the valence densities of down and up quarks, respectively, in the proton. The quantities $s(x)$ and $c(x)$ represent the distributions of strange and charm quarks, while $\bar{c}(x)$ is the distribution of charm antiquarks.

When experimental data is reported by collaborations such as CCFR~\cite{Seligman:1997mc}, NuTeV data~\cite{NuTeV:2005wsg} and CHORUS data~\cite{CHORUS:2005cpn}.

This expression characterizes the momentum density of partons, including valence quarks and antiquarks, within the nucleon and provides essential insights into the quark-gluon dynamics in the proton. 

For the current analysis, we employ the following standard parameterizations for the valence distributions, $xu_v$ and $xd_v$, using Gegenbauer polynomials:

\begin{equation}\label{eq:xuvQ0}
	xu_v = {\cal N}_u x^{\alpha_{u_v}}(1-x)^{\beta_{u_v}}\left(1+\sum_{i=1}^{3}a_u^iC^{\frac{7}{2}}(i,1-2x)\right)\,,
\end{equation}

\begin{eqnarray} \label{eq:xdvQ0}
	xd_v &=& \frac{{\cal N}_d} {{\cal N}_u}(1 - x)^{\beta_{d_v}} xu_v\,.\nonumber\\
	xd_v &=& {\cal N}_d x^{\alpha_{u_v}}(1-x)^{\beta_{u_v}+\beta_{d_v}}\left(1+\sum_{i=1}^{3}a_u^iC^{\frac{7}{2}}(i,1-2x)\right)\,,\nonumber\\
\end{eqnarray}

where $Q^2_0 =1 GeV^2$ is the input scale and the $C^{\frac{7}{2}}(i,1-2x)$ are Gegenbauer polynomials.  The normalizations
${{\cal N}}_u$ and ${{\cal N}}_d$ are being fixed by $\int_0^1 u_v
dx=2$ and $\int_0^1 d_v dx=1$, respectively.

\begin{figure}
 	\begin{center}
 		\includegraphics[clip,width=0.5\textwidth]{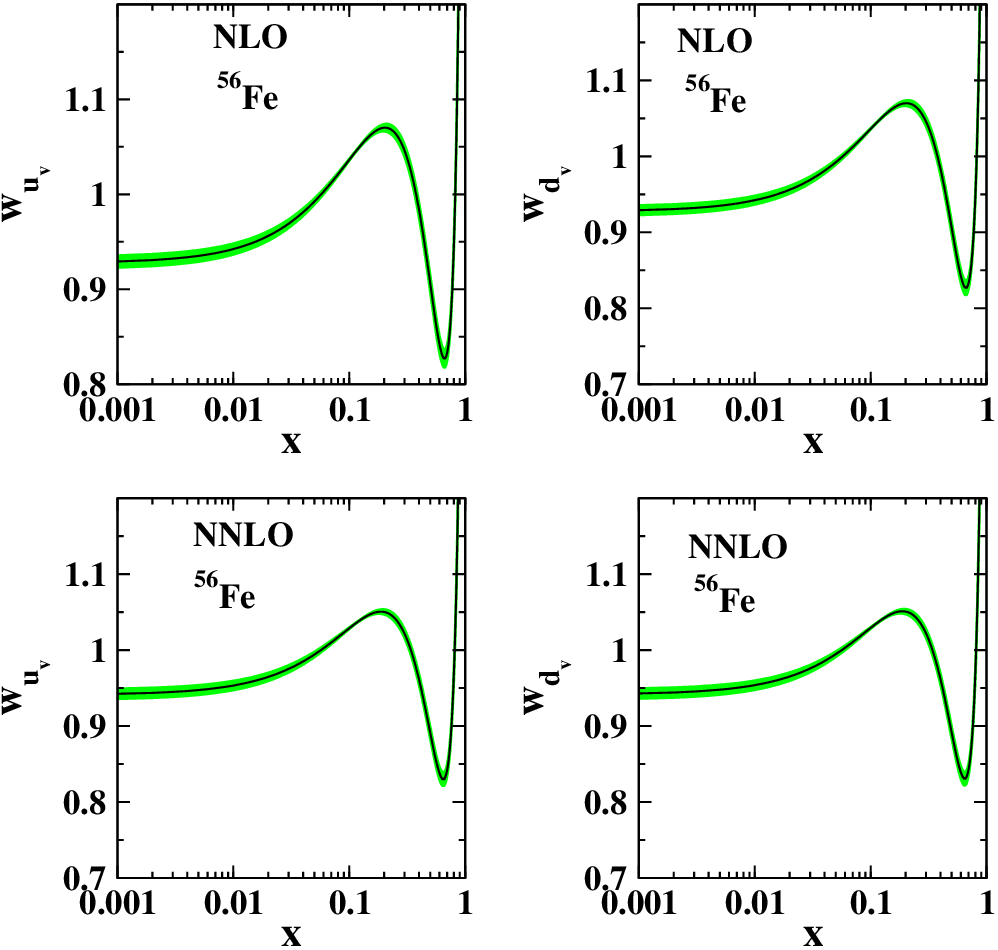}% 
 		%\vspace{-5 cm}
 		\caption{\sf The nuclear weight function calculated in refs.~\cite{AtashbarTehrani:2012xh,Khanpour:2016pph} in NLO and NNLO approximation for $^{56}Fe$. } \label{fig1}
 	\end{center}
 \end{figure}

\begin{figure}
 	\begin{center}
 		\includegraphics[clip,width=0.5\textwidth]{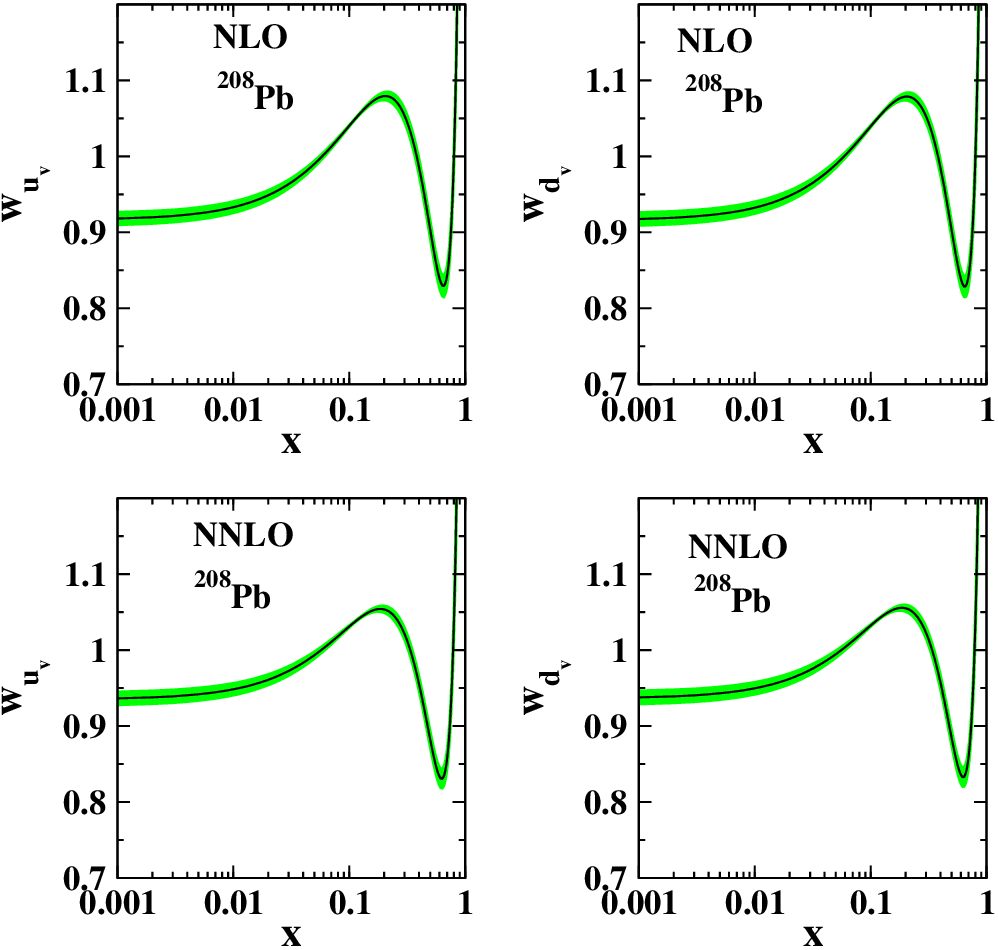}% 
 		%\vspace{-5 cm}
 		\caption{\sf The nuclear weight function calculated in the refs.~\cite{AtashbarTehrani:2012xh,Khanpour:2016pph} in NLO and NNLO approximation for $^{208}Pb$. } \label{fig2}
 	\end{center}
 \end{figure}

We use the neutrino-nucleus data so we take into account the nuclear effects and use the nuclear weight function in the structure function $xF_3$ calculations. We select the new form for $xu_v$ and $xd_v$ that follows the following expression
\begin{eqnarray}
	xu_v^A&=&{\cal W}_{u_v}\frac{Z xu_v+N xd_v}{A} \nonumber\\
	xd_v^A&=&{\cal W}_{d_v}\frac{N xu_v+Z xd_v}{A}
\end{eqnarray}
The shape of the weight function is as follows:
\begin{eqnarray}
{\cal W}_{u_v}&=&1+\left(1-\frac{1}{A^{1/3}}\right)\frac{A_u+c_1x+c_2x^2+c_3x^3}{(1-x)^{0.4}}\nonumber\\
{\cal W}_{d_v}&=&1+\left(1-\frac{1}{A^{1/3}}\right)\frac{A_d+c_1x+c_2x^2+c_3x^3}{(1-x)^{0.4}}
\end{eqnarray}
This distribution of nuclear weight was calculated in the refs. \cite{AtashbarTehrani:2012xh,Khanpour:2016pph} in the NLO and NNLO approximation and are shown in Figs.~\ref{fig1} and \ref{fig2}. In our analysis, we used the NNLO weight function for the NNNLO approximation, because no group has obtained a weight function in this approximation, i.e., NNNLO, for the nuclear structure functions.
The deep inelastic scattering data of the neutrino nucleus are for iron and lead, and we selected A=56, Z=26 and N=A-Z to represent the number of neutrons for iron and A=208, Z=82 for lead.
 

 The Mellin transform of these functions are defined as
\begin{eqnarray}
	u_v^A(N,Q_0^2)&=&\int_0^1xu_v^A(x,Q_0^2)x^{n-2}dx\nonumber\\
	d_v^A(N,Q_0^2)&=&\int_0^1xd_v^A(x,Q_0^2)x^{n-2}dx.
\end{eqnarray}
The evolution equation of the nonsinglet structure function
$xF_3(x,Q^2)$ in Mellin space, extended to the NLO-loop order, can be found in the reference~\cite{Gluck:1989ze}.

\begin{eqnarray}
	F_3(N,Q^2)&=&\left(1+a\;C_{3}^{(1)}(N)\right)\nonumber\\&&\times F_3(N,Q_0^2)
	\left(\frac{a}{a_0}\right)^{-\hat{P}_0(N)/{\beta_0}} \nonumber
	\\
	&&\Biggl\{1 - \frac{1}{\beta_0} (a - a_0) \left[\hat{P}_1^+(N)
	- \frac{\beta_1}{\beta_0} \hat{P}_0(N) \right]\Bigg\}~ \nonumber\\
\end{eqnarray}
The evolution equation of $xF_3(x,Q^2)$, extended to the N$^2$LO-loop order, can be found in the reference \cite{Blumlein:2006be,Blumlein:2021lmf}. Within this framework, the nonsinglet structure functions can be expressed as follows:

\begin{eqnarray}
	F_3(N,Q^2)&=&\left(1+a\;C_{3}^{(1)}(N)+a^2\;C_{3}^{(2)}(N)\right)\nonumber\\&&\times F_3(N,Q_0^2)
	\left(\frac{a}{a_0}\right)^{-\hat{P}_0(N)/{\beta_0}} \nonumber
	\\
	&&\Biggl\{1 - \frac{1}{\beta_0} (a - a_0) \left[\hat{P}_1^+(N)
	- \frac{\beta_1}{\beta_0} \hat{P}_0(N) \right] \nonumber\\
	& & - \frac{1}{2 \beta_0}\left(a^2 - a_0^2\right)
	\left[\hat{P}_2^+(N) - \frac{\beta_1}{\beta_0}
	\hat{P}_1^+(N)\right. \nonumber \\ &&\left.+ \left(
	\frac{\beta_1^2}{\beta_0^2} -
	\frac{\beta_2}{\beta_0} \right) \hat{P}_0(N)   \right] \nonumber\\
	& & + \frac{1}{2 \beta_0^2} \left(a - a_0\right)^2
	\left(\hat{P}_1^+(N) - \frac{\beta_1}{\beta_0} \hat{P}_0(N)
	\right)^2 \Bigg\}~.\nonumber\\
\end{eqnarray}

The NLO and N$^2$LO Wilson coefficient functions $C_3^{(1)}$ and $C_3^{(2)}$ in Mellin $N$ space can be determined easily using the references \cite{Moch:1999eb, vanNeerven:1999ca}. The splitting functions in Mellin $N$ space can be found in Refs. \cite{Floratos:1981hs, Moch:2004pa, Vogt:2004mw, JimenezDelgado:2008rdu,Moch:2017uml}.

The expansion coefficients $\beta_k$ of the $\beta$-function of QCD are known up to $k=3$, corresponding to the N$^3$LO (next-to-next-to-next-to-leading order) \cite{Tarasov:1980au,Larin:1993tp,Blumlein:2021enk}. These $\beta_k$ coefficients are important for determining the evolution of the strong coupling constant $\alpha_s$ with the scale $Q^2$ and are significant in the perturbative calculations of various QCD processes:
\begin{eqnarray}
	\label{beta-exp}
	\beta_0 &=& 11-2/3~n_f\;,
	\nonumber \\
	\beta_1 &=& 102-38/3~n_f\;,
	\nonumber \\
	\beta_2 &=& 2857/2-5033/18~n_f+325/54~n_f^2\;,
	\nonumber \\
	\beta_3 &=& 29243.0 - \: 6946.30~n_f + 405.089~n_f^2
	+ 1093/729~n_f^3~,\nonumber\\
\end{eqnarray}
here $n_f$ stands for the number of effectively massless quark
flavors and $\beta_k$ denote the coefficients of the usual
four-dimensional $\overline{MS}$ beta function of QCD.

 Within this framework, the nonsinglet structure functions can be expressed as follows in N$^3$LO:
 \begin{widetext}
\begin{eqnarray}
	F_3(N,Q^2)&=&\left(1+a_s\;C_{3,{\rm
			NS}}^{(1)}(N)+a_s^2\;C_{3,{\rm NS}}^{(2)}(N)+a_s^3\;C_{3,{\rm
			NS}}^{(3)}(N)\right) F_3(N,Q_0^2)
	\nonumber
	\\
	&&\times\left(\frac{a_s}{a_0}\right)^{-\hat{P}_0(N)/{\beta_0}}\Biggl\{1
	- \frac{1}{\beta_0} (a_s - a_0) \left[\hat{P}_1^+(N)
	- \frac{\beta_1}{\beta_0} \hat{P}_0(N) \right] \nonumber\\
	& & - \frac{1}{2 \beta_0}\left(a_s^2 - a_0^2\right)
	\left[\hat{P}_2^+(N) - \frac{\beta_1}{\beta_0} \hat{P}_1^+(N) +
	\left( \frac{\beta_1^2}{\beta_0^2} - \frac{\beta_2}{\beta_0}
	\right) \hat{P}_0(N)   \right] \nonumber\\ & & + \frac{1}{2
		\beta_0^2} \left(a_s - a_0\right)^2 \left(\hat{P}_1^+(N) -
	\frac{\beta_1}{\beta_0} \hat{P}_0(N) \right)^2 \nonumber\\ & &
	%---
	- \frac{1}{3 \beta_0} \left(a_s^3 - a_0^3\right)
	\Biggl[\hat{P}_3^+(N) - \frac{\beta_1}{\beta_0} \hat{P}_2^+(N) +
	\left(\frac{\beta_1^2}{\beta_0^2} - \frac{\beta_2}{\beta_0}\right)
	\hat{P}_1^+(N) \nonumber\\ & & +\left(\frac{\beta_1^3}{\beta_0^3}
	-2 \frac{\beta_1 \beta_2}{\beta_0^2} + \frac{\beta_3}{\beta_0}
	\right) \hat{P}_0(N)  \Biggr] \nonumber\\ & & + \frac{1}{2
		\beta_0^2} \left(a_s-a_0\right)\left(a_0^2 - a_s^2\right)
	\left(\hat{P}_1^+(N)-\frac{\beta_1}{\beta_0} \hat{P}_0(N) \right)
	\nonumber\\ & & \times \left[\hat{P}_2(N) -
	\frac{\beta_1}{\beta_0} \hat{P}_1(N) -
	\left(\frac{\beta_1^2}{\beta_0^2} - \frac{\beta_2}{\beta_0}
	\right) \hat{P}_0(N)
	\right]
	\nonumber\\ && - \frac{1}{6 \beta_0^3} \left(a_s-a_0\right)^3
	\left(\hat{P}_1^+(N)-\frac{\beta_1}{\beta_0} \hat{P}_0(N)
	\right)^3 \Bigg\}~.
\end{eqnarray}
\end{widetext}
and
\begin{eqnarray}
	F_3(N,Q_0^2)=u_v^A(N,Q_0^2)+d_v^A(N,Q_0^2)
\end{eqnarray}
Here $a_s(=\alpha_s/4\pi)$ and $a_0$ represent the strong coupling constant at the scales of $Q^2$ and $Q_0^2$ respectively. $C_{3,NS}^{(m)}(N)$ refers to the nonsinglet Wilson coefficients in ${\it{O}}(a_s^m)$, which can be found in the cited reference \cite{Blumlein:2022gpp}. The term "$\hat{P}_m$" also denotes the Mellin transforms of the $(m+1)$-loop splitting functions.

The strong coupling constant $a_{s}$ is of utmost significance in the present paper regarding the evolution of parton densities. At $N^{m}LO$, the scale dependence of $a_{s}$ is determined by
\begin{eqnarray}
	\label{as-eqn}
	\frac{d\, a_{s}}{d \ln Q^2} \; = \; \beta_{N^mLO}(a_{s})
	\; = \; - \sum_{k=0}^m \, a_{s}^{k+2} \,\beta_k \;.
\end{eqnarray}
 In complete
N$^3$LO-loop approximation and using the $\Lambda$-parametrization, the
running coupling is given by \cite{Vogt:2004ns,Chetyrkin:1997sg}:
\begin{eqnarray}
	\label{as-exp1}
	a_s(Q^2)
	&&=\frac{1}{\beta_0{L_{\Lambda}}}  -
	\frac{1}{(\beta_0{L_{\Lambda}})^2}~b_1 \ln {L_{\Lambda}}+
	\frac{1}{(\beta_0{L_{\Lambda}})^3} \nonumber\\&& \left[b_1^2 \left(\ln^2 {L_{\Lambda}}-\ln {L_{\Lambda}}-1
	\right) + b_2\right]+\frac{1}{(\beta_0{L_{\Lambda}})^4} \nonumber\\
	&&\left[b_1^3 \left(-\ln^3 {L_{\Lambda}} +
	\frac{5}{2} \ln^2{L_{\Lambda}} +2 \ln{L_{\Lambda}}-\frac{1}{2}\right)\right.\nonumber\\
	&&-\left. 3b_1 b_2\ln{L_{\Lambda}}
	+\frac{b_3}{2} \,\right]~,
\end{eqnarray}
where $L_{\Lambda}\equiv ln (Q^2/\Lambda^2)$, $b_k\equiv \beta_k
/\beta_0$, and $\Lambda$ is the QCD scale parameter. The first
line of Eq.~(\ref{as-exp1}) includes the  the NLO-loop
coefficients, the second line is the N$^2$LO-loop and the third line
denotes the N$^3$LO-loop correction. Equation~(\ref{as-exp1}) solves the
evolution equation (\ref{as-eqn}) only up to higher orders in
$1/L_{\Lambda}$. The functional form of $\alpha_s(Q^2)$, in N$^3$LO-loop
approximation and for 6 different values of $\Lambda$.  To be able to compare
with other measurements of $\Lambda$ we adopt the matching of
flavor thresholds at $Q^2=m_c^2$ and $Q^2=m_b^2$ with $m_c=1.5$
GeV and $m_b=4.5$ GeV as described in \cite{BAR,R1998}.

\section{Jacobi polynomials and the procedure of QCD fits}\label{sec:sec4}

One of the simplest and fastest methods for reconstructing the structure function from QCD predictions for its Mellin moments is through the expansion of Jacobi polynomials. The Jacobi polynomials
are especially suitable for this purpose since they allow one to
factor out an essential part of the $x$-dependence of  structure
function into the weight function \cite{Parisi:1978jv}.

According to this method, one can relate the $xF_3$ structure
function with its Mellin moments
\begin{eqnarray} xF_{3}^{N_{max}}(x,Q^2)&=&x^{\beta}(1-x)^{\alpha}
	\sum_{n=0}^{N_{max}}\Theta_n ^{\alpha,
		\beta}(x)\nonumber\\&&\sum_{j=0}^{n}c_{j}^{(n)}{(\alpha ,\beta )}
	F_{3}(j+2,Q^2), \label{eg1Jacob} \nonumber\\
	\end{eqnarray}
 where $N_{max}$ is the number of polynomials.  Jacobi polynomials of order $n$
\cite{Parisi:1978jv}, $\Theta_n ^{\alpha, \beta}(x)$, satisfy the
orthogonality condition with the weight function $w^{\alpha
	\beta}=x^{\beta}(1-x)^{\alpha}$
\begin{equation}
	\int_{0}^{1}dx\;w^{\alpha \beta} \Theta_{k} ^{\alpha , \beta}(x)
	\Theta_{l} ^{\alpha , \beta}(x)=\delta_{k,l}\ .\label{e8}
\end{equation}
In the above, $c_{j}^{(n)}{(\alpha ,\beta )}$ are the coefficients
expressed through $\Gamma$-functions and satisfying the
orthogonality relation in Eq.~(\ref{e8}) and $F_{3}(j+2,Q^2)$ are
the moments determined in the previous section. $N_{max}$,
$\alpha$ and $\beta$  have to be chosen so as to achieve the
fastest convergence of the series on the right-hand side of
Eq.~(\ref{eg1Jacob}) and to reconstruct $F_2$ with the required
accuracy. In our analysis we use $N_{max}=9$, $\alpha=3.0$ and
$\beta=0.5$. The same method has been applied to calculate the
nonsinglet structure function $xF_3$ from their moments
\cite{Kataev:1997nc,Kataev:1998ce, Kataev:1999bp,Kataev:2001kk}
and for polarized structure function $xg_1$ \cite{Nematollahi:2023dvj,Mirjalili:2022cal,Nematollahi:2021ynm,Salajegheh:2018hfs}. Obviously the
$Q^2$-dependence of the polarized structure function is defined by
the $Q^2$-dependence of the moments.
The evolution equations allow for the calculation of the $Q^2$-dependence of parton distributions provided at a certain reference point, $ Q_0^2 $. These distributions are typically parameterized based on plausible theoretical assumptions regarding their behavior near the endpoints, $ x = 0 $ and $ x = 1 $.
%\section{The  Procedure of the QCD Fits of  $xF_3$ Data}\label{sec:sec5}
For the data utilized in the global analysis, most experiments combine various systematic errors into one effective error for each data point, along with the statistical error. Additionally, the fully correlated normalization error of the experiment is usually specified separately. Therefore, it is natural to adopt the following definition for the effective $\chi^2$ \cite{Stump:2001gu}.
\begin{eqnarray}
	\chi _{\mathrm{global}}^{2} &=&
	\sum_{n} w_{n} \chi _{n}^{2}\;,\; (n\;%
	\mbox{labels the different experiments})\nonumber
	\label{eq:Chi2global}
	\\
	\chi _{n}^{2} &=&\left(\frac{1-{\cal N}_{n}}{\Delta{\cal
			N}_{n}}\right)^{2} +\sum_{i}\left( \frac{{\cal
			N}_{n}xF_{3,i}^{data}-xF_{3,i}^{theor}}{{\cal N}_{n}\Delta
		xF_{3,i}^{data}} \right)^{2}\;. \label{eq:Chi2n}\nonumber\\
\end{eqnarray}

For the $n^{\mathrm{th}}$ experiment, $xF_{3,i}^{data}$, $\Delta
xF_{3,i}^{data}$, and $xF_{3,i}^{theor}$ denote the data value, measurement uncertainty
(statistical and systematic combined) and theoretical value for
the $i^{\mathrm{th}}$ data point. ${\Delta{\cal N}_{n}}$ is the
experimental normalization uncertainty and ${\cal N}_{n}$ is an
overall normalization factor for the data of experiment $n$. The
factor $w_{n}$ is a possible weighting factor (with default value
1).  However, we allowed for a relative normalization shift ${\cal	N}_{n}$ between the different data sets within the normalization
uncertainties ${\Delta{\cal N}_{n}}$ quoted by the experiments.

Now the sums in $\chi _{\mathrm{global}}^{2}$ run over all data
sets and in each data set over all data points. The minimization
of the  above $\chi^2$ value to determine the best parametrization
of the unpolarized parton distributions is done using the program
{\tt MINUIT} \cite{MINUIT}.

\section{Fit Results}\label{sec:sec6}
%--------------------------------
The data for the charged-current structure functions $xF_3(x, Q^2)$ used in our analysis are listed in Table.~\ref{table:xf3data}. The $x$ and $Q^2$ ranges, the number of data points and the related references are also listed in this table.
\begin{table*}[htb]
	\begin{tabular}{|c|c|c|c|c|}
		\hline
		Experiment & $x$    & Q$^2$ & Number of data points  & Reference        \\      \hline   \hline
		CCFR	 & $0.0075 \leq x \leq 0.75$    & $1.3 \leq Q^2 \leq 125.9 $ & 116 & \cite{Seligman:1997mc}  \\
		NuTeV	 &   $0.015 \leq x \leq 0.75$  &$3.162 \leq Q^2 \leq 50.118$  & 64 & \cite{NuTeV:2005wsg}    \\
		CHORUS	 &  $0.02 \leq x \leq 0.65$   &$2.052 \leq Q^2 \leq 81.55$  & 50 & \cite{CHORUS:2005cpn}     \\   \hline
	\end{tabular}
	\caption{ Published data points for charged-current structure functions $xF_3(x, Q^2)$ used in the present global fit. The $x$ and $Q^2$ ranges, the number of data points and the related references are also listed. \label{table:xf3data} }
\end{table*}
%--------------------------------
The CCFR~\cite{Seligman:1997mc} and NuTeV~\cite{NuTeV:2005wsg} collaborations at Fermilab conduct neutrino deep inelastic scattering experiments using an iron target, which are subsequently adjusted to account for an isoscalar target. They cover much of the same kinematic range of momentum fraction $x$, but CCFR covers slightly higher $Q^2$. At high values of $x$, the predictions are mainly determined by the valence up quark distribution, which is very well constrained by the charged-current DIS structure function data. We also include recent data from the CHORUS~\cite{CHORUS:2005cpn} collaboration, which are taken from a lead target and cover a similar range in $x$ compared with CCFR. The NuTeV data seem to be more precise. In practice, we find the high-$x$ NuTeV and CHORUS data very difficult to fit, leading to higher values of $\chi^2$.

\begin{table*}
	\caption{ Best fit parameters and uncertainties of the
		Mellin fits at NLO , N$^2$LO and N$^3$LO at the initial scale
		$Q_0^2 = 1.0$~GeV$^2$. }
	{ \begin{tabular}{cccc}
			\hline
			& \multicolumn{3}{c}{Mellin}  \\ \hline
			& NLO & N$^2$LO & N$^3$LO \\ \hline
			$a_{u}^{1}$ & \multicolumn{1}{l}{$-0.20077\pm0.0013422$} & \multicolumn{1}{l}{%
				$ 0.026755\pm 0.00332961$} &$-0.16898\pm0.022263$\\ 
			$a_{u}^{2}$ & \multicolumn{1}{l}{$0.030599\pm 0.000367904$} & \multicolumn{1}{l}{$%
				-0.0195933\pm 0.00170209$} &$0.0235584\pm0.00059855$ \\ 
			$a_{u}^{3}$ & \multicolumn{1}{l}{$-0.0033036\pm 0.00010995$} & \multicolumn{1}{l}{%
				$ 0.003302035\pm 0.000536726$} &$-0.0020921\pm0.00017266$ \\ 
			$\alpha _{u_{v}}$ & \multicolumn{1}{l}{$0.63134\pm 0.008237101$} & 
			\multicolumn{1}{l}{$0.782549\pm 0.00586995$} &$0.78679\pm0.016633$\\ 
			$\beta _{u_{v}}$ & \multicolumn{1}{l}{$4.05839\pm0.055065$} & 
			\multicolumn{1}{l}{$2.962457\pm 0.0272975$} &$4.24605\pm0.057478$\\ 
			$\beta _{d_{v}}$ & $-0.0075767\pm 0.16530$ & $0.00262071\pm 0.082775$ &$0.00105091\pm0.17327$ \\ 
			$\alpha _{s}(Q_{0}^{2})$ & $0.49941\pm0.017535$ & $0.45056\pm0.007143$  &$0.41758\pm0.01003$\\ 
			$\alpha _{s}(M_{z}^{2})$ & $0.12101\pm0.000551$ & $0.11958\pm0.007143$&$0.118011\pm0.000505$\\ \hline
			$\chi ^{2}/d.o.f$ & $368.8118/223=1.6538$ & $327.6235/223= 1.4691$ &$310.0293/223=1.39026$ \\ \hline
		\end{tabular} \label{ta2}}%
\end{table*}

\begin{figure}
	\begin{center}
		\includegraphics[clip,width=0.5\textwidth]{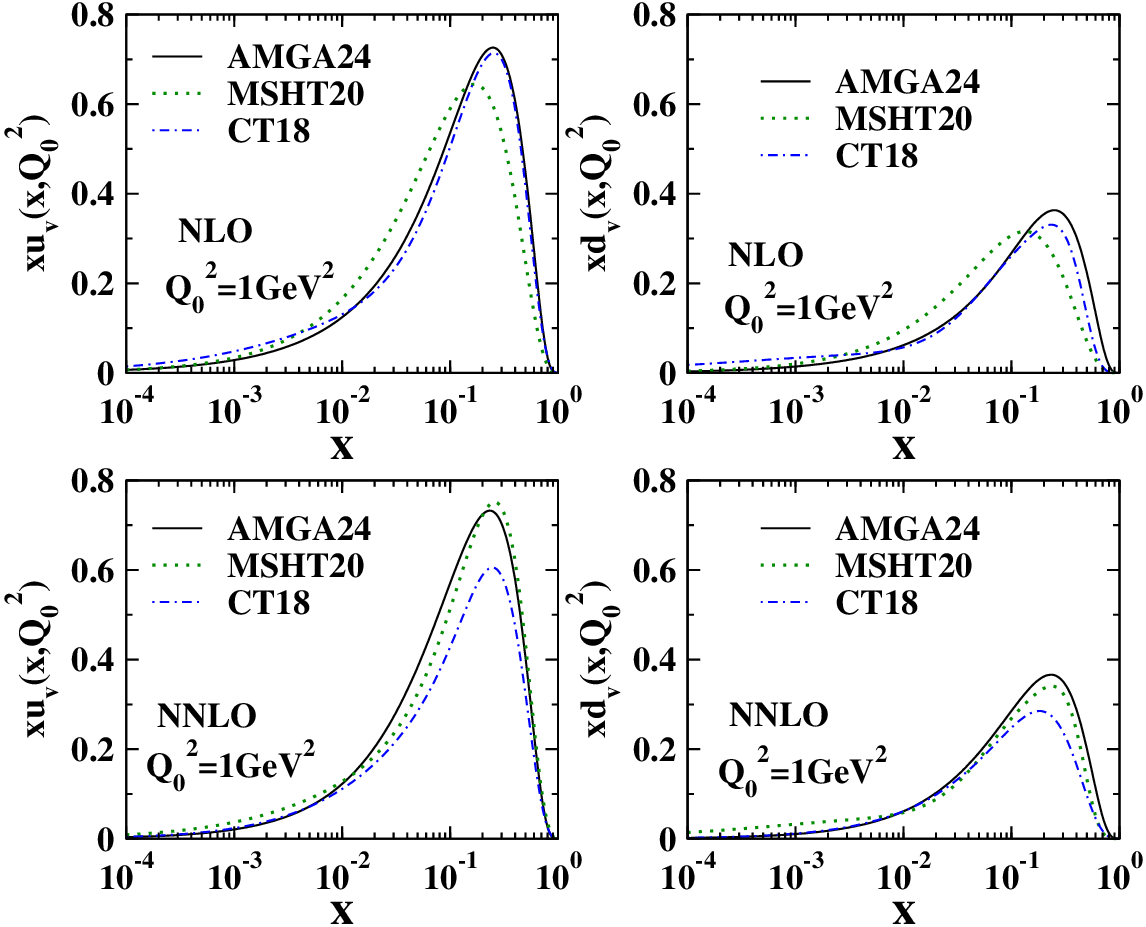}% 
		%\vspace{-5 cm}
	\caption{\sf $xu_v$ and $xd_v$ in NLO and N$^2$LO approximation in Mellin model compare with MSHT20~\cite{Bailey:2020ooq} and CT18~\cite{Hou:2019efy} } \label{fig3} 
	\end{center}
\end{figure}

\begin{figure}
	\begin{center}
		\includegraphics[clip,width=0.5\textwidth]{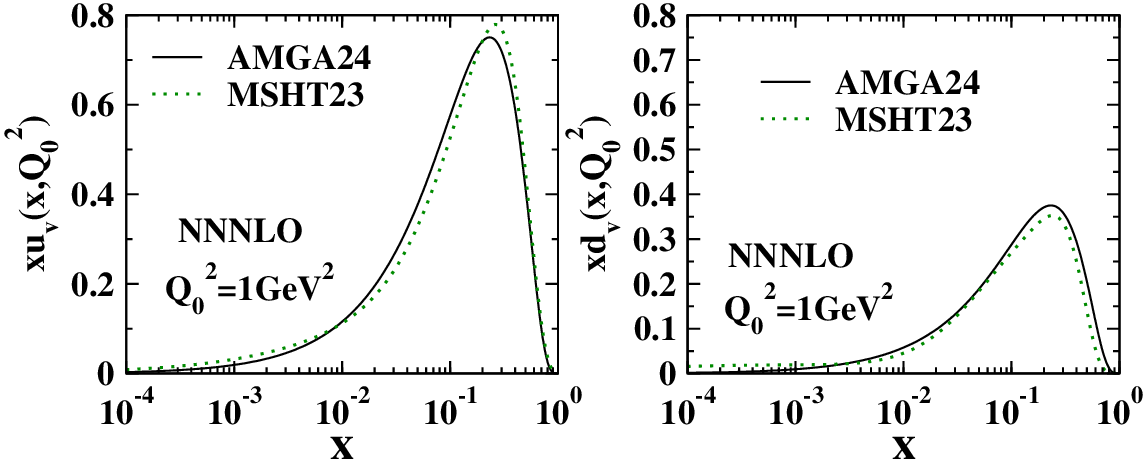}% 
		%\vspace{-5 cm}
		\caption{\sf $xu_v$ and $xd_v$ in N$^3$LO approximation in Mellin model compare with MSHT23~\cite{McGowan:2022nag}  } \label{fig4} 
	\end{center}
\end{figure}

\begin{figure}
	\begin{center}
		\includegraphics[clip,width=0.5\textwidth]{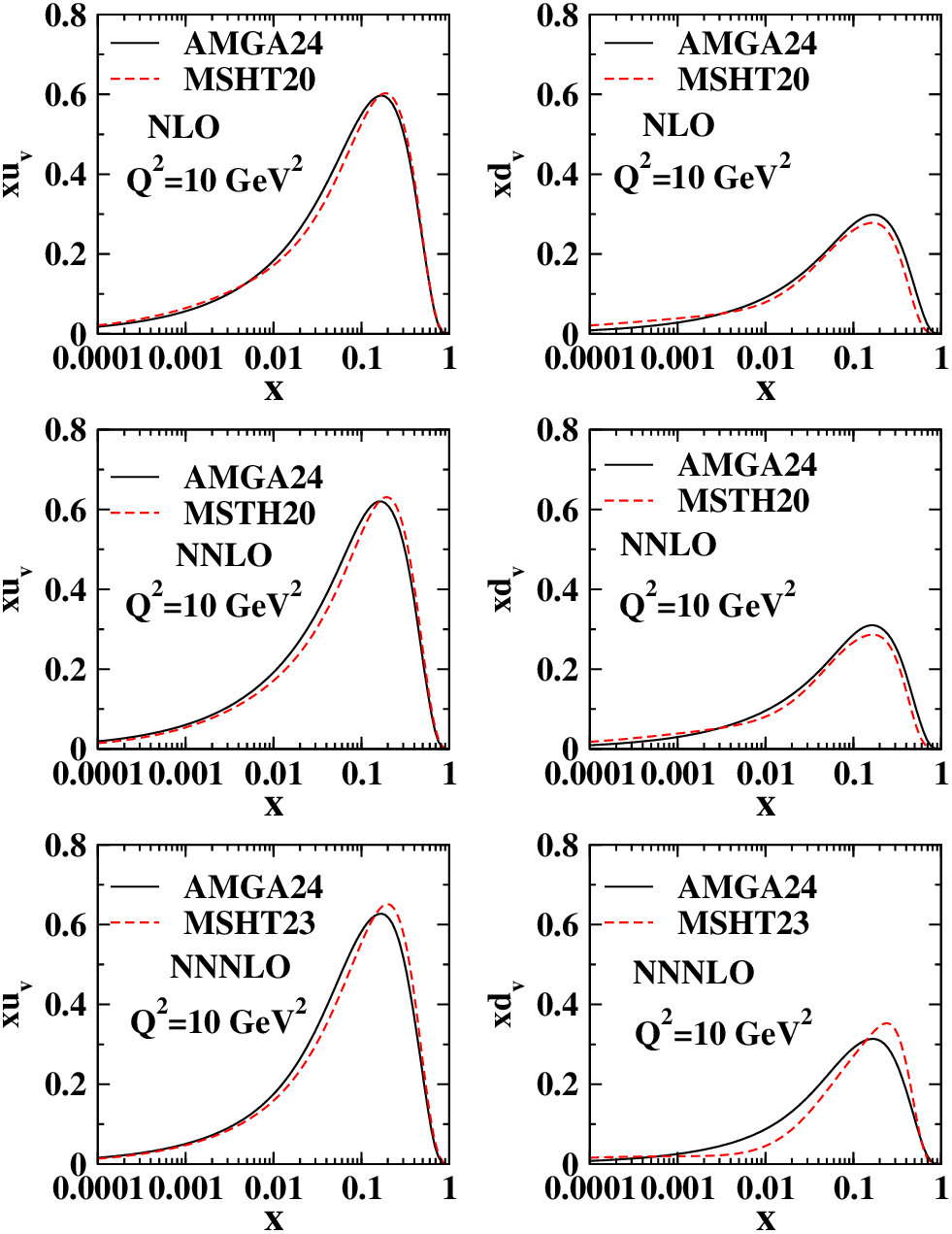}% 
		%\vspace{-5 cm}
		\caption{\sf $xu_v$ and $xd_v$ at $Q^2=10 GeV^2$ in NLO, N$^2$LO and N$^3$LO approximation in Mellin model compare with MSHT20~\cite{Bailey:2020ooq} and MSHT23~\cite{McGowan:2022nag}  } \label{fig5} 
	\end{center}
\end{figure}

\begin{figure}
	\begin{center}
		\includegraphics[clip,width=0.5\textwidth]{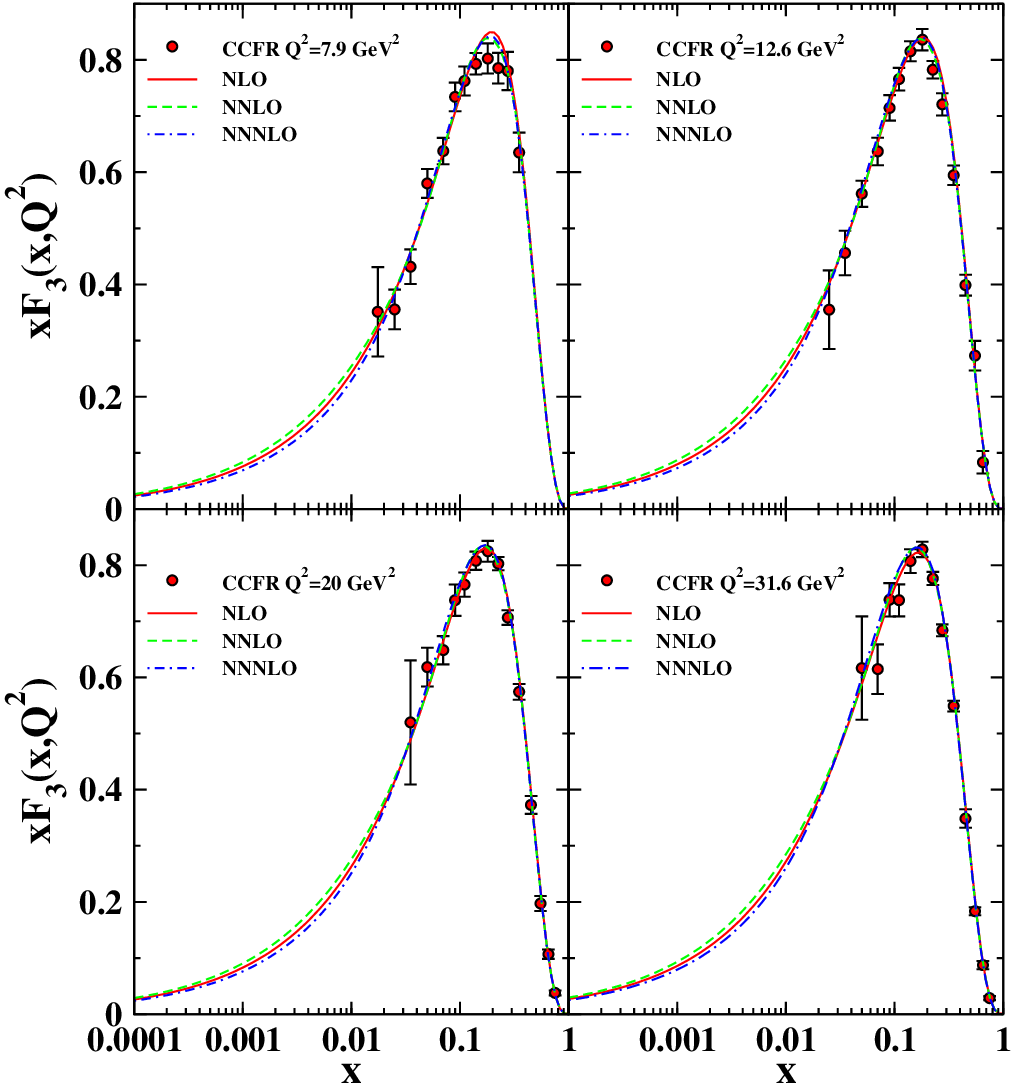}% 
		%\vspace{-4 cm}
		\caption{\sf $xF_3$  in NLO , N$^2$LO and N$^3$LO approximation in Mellin  model compare with {\tt CCFR} data~\cite{Seligman:1997mc} }\label{fig6}
	\end{center}
\end{figure}

\begin{figure}
	\begin{center}
		\includegraphics[clip,width=0.5\textwidth]{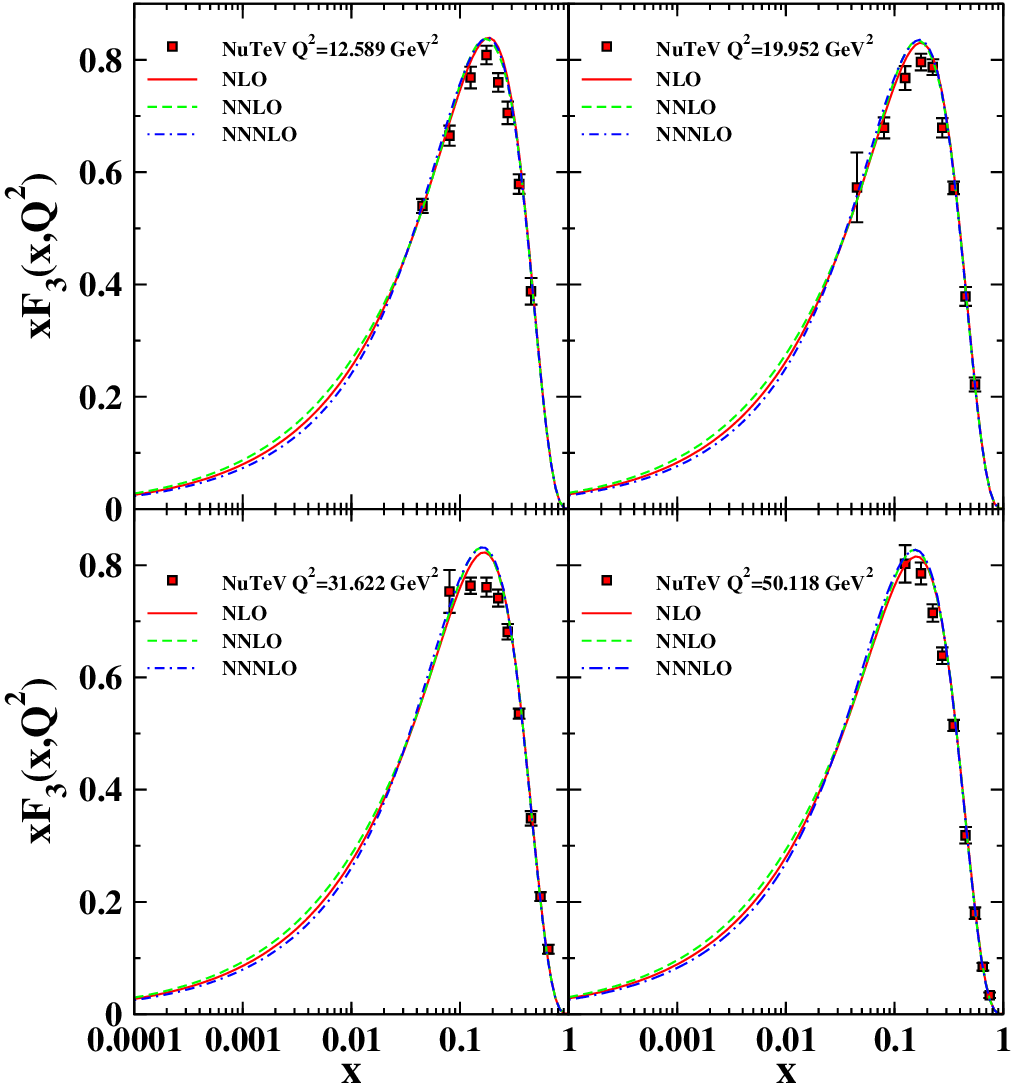}%
		%\vspace{-0.4 cm}
		\caption{\sf $xF_3$ at NLO , N$^2$LO and N$^3$LO  approximation in Mellin  model compare with {\tt NuTeV} data~\cite{NuTeV:2005wsg}.} \label{fig7} 
	\end{center}
\end{figure}

\begin{figure} 
	\begin{center}
		\includegraphics[clip,width=0.5\textwidth]{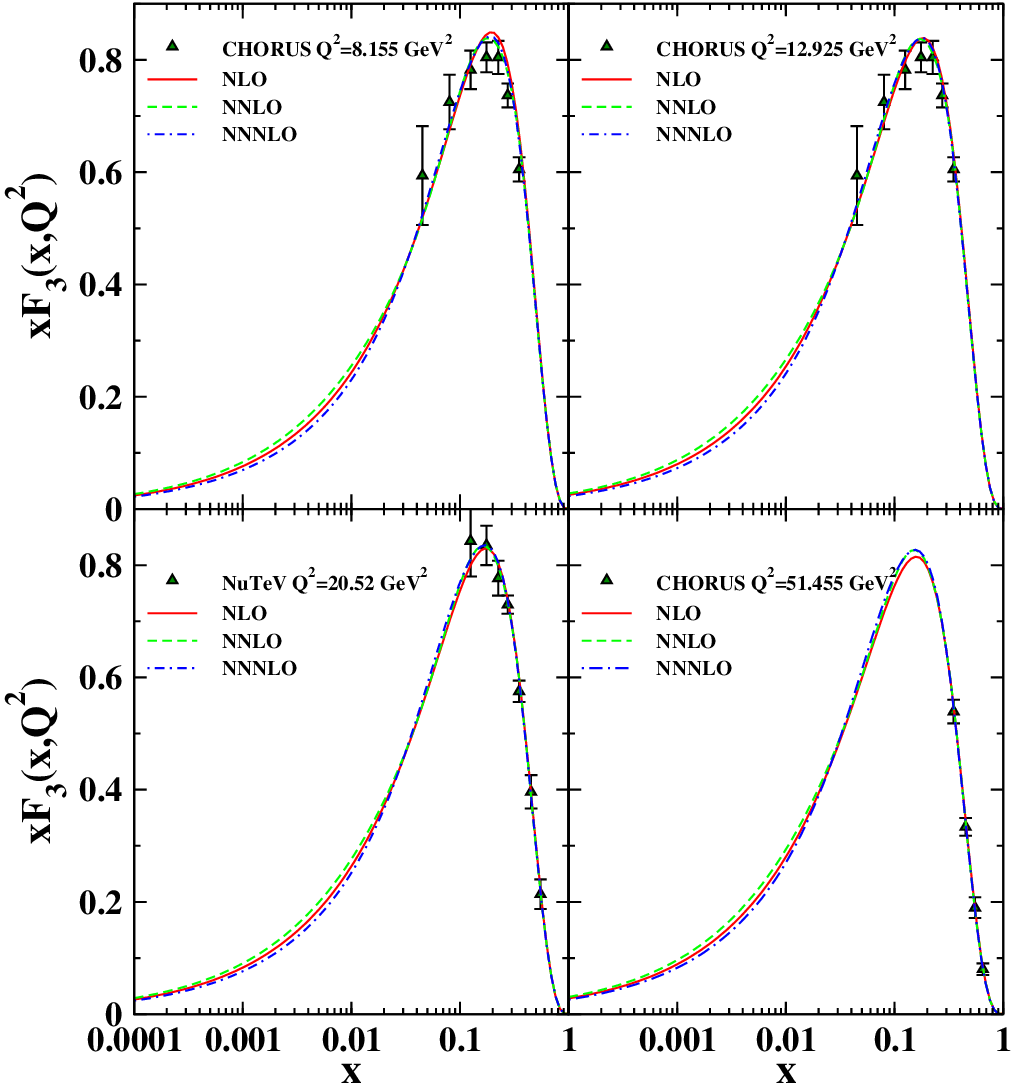}%
		%\vspace{-4 cm}
		\caption{\sf $xF_3$ at NLO , N$^2$LO and N$^3$LO approximation in Mellin  model compare with {\tt CHORUS} data~\cite{CHORUS:2005cpn}.} \label{fig8}
	\end{center}
\end{figure}

The accuracy of this result stems from the obtained $\chi^2/d.o.f$ value by fitting the initial valence PDFs at $Q_0^2 $= 1 ${GeV}^2$, as detailed in Table~\ref{ta2}. The world average value for $\alpha _{s}(M_{z}^{2}) = 0.1179\pm8.5\times10^{-6}$ , as reported in Ref.~\cite{ParticleDataGroup:2020ssz} is in good agreement with the reported ones in Table ~\ref{ta2}.\\
In Figs.~\ref{fig3}, \ref{fig4} and~\ref{fig5}, valence PDFs for different approaches at $Q^2 = 1 \, \text{GeV}^2$ and $Q^2 = 10 \, \text{GeV}^2$ have been depicted in NLO, NNLO, and NNNLO approximations. As can be seen, the agreement with the results of MSHT20~\cite{Bailey:2020ooq} improves notably at NNLO. The results of the analysis by MSTH23~\cite{McGowan:2022nag} in N$^3$LO are very close to those of AMGA24 at $Q^2 = 1 \, \text{GeV}^2$ .\\
Additionally, the results for the $xF_3$ proton structure function in NLO, NNLO and NNNLO approximations are presented in Fig.~\ref{fig6}. The comparison to experimental data from the CCFR collaboration~\cite{Seligman:1997mc}, which provides direct measurements of the $xF_3$ structure function, is also shown. The agreement between the theoretical predictions and the experimental data highlights the reliability of the current theoretical framework. \\
Furthermore, in Figs.~\ref{fig7} and~\ref{fig8}, the $xF_3$ structure functions at NLO, NNLO, and NNNLO are compared with experimental data from the NuTeV~\cite{NuTeV:2005wsg} and CHORUS~\cite{CHORUS:2005cpn} collaborations at various values of $Q^2$. These comparisons further validate the theoretical predictions and provide additional constraints on the PDFs.
In Ref.~\cite{Tooran:2019cfz}, the $ \chi^2 $ for NNLO was reported as 1.482, while in the current model, we obtain 1.4691, indicating that the choice of Gegenbauer polynomial is appropriate.  We have done calculations up to NNNLO order accurately.
As can be observed, our analysis derives the exact solution by incorporating the splitting function as per Ref.~\cite{Moch:2017uml} and the Wilson coefficient following Ref.~\cite{Blumlein:2022gpp} within the NNNLO approximation, unlike Ref.~\cite{Kataev:1998ce} which employed the Pad\'e approximation method for the $ xF_3 $ structure function at NNNLO. We have established that the Gegenbauer polynomial is as effective as the Chebyshev polynomial, corroborated by findings from the MSHT20 and MSTH23 groups in Refs.~\cite{Bailey:2020ooq} and~\cite{McGowan:2022nag}, respectively. Our analysis successfully computes the non-singlet component of the structure function using a minimal dataset and a streamlined approach, achieving congruence with other established models. 

\section{  the Gross-Llewellyn Smith (GLS) sum rule }\label{sec:sec7}
Another intriguing issue revolves around the extraction of the value of the Gross-Llewellyn Smith (GLS) sum rule. The GLS sum rule is a crucial property in the context of deep inelastic neutrino-nucleon scattering. In the framework of the quark-parton model, the GLS sum rule, which is associated with the $xF_3$  structure function, is expressed as~\cite{Gross:1969jf}.

The GLS sum rule relates the integral of the $xF_3(x,Q^2)$  structure function over the entire $x$:
%-------------------------------------------
\begin{equation}\label{eq:GLSSL} 
	\text{ GLS} (Q^2) = \frac{1}{2} \int_0^1 \frac{xF_3^{\bar{\nu} p + \nu p}(x,Q^2)}{x} dx \,.
\end{equation}
%-------------------------------------------
Its experimental verification provides valuable constraints on the parameters of the electroweak theory and is crucial for understanding the interplay between weak and strong interactions in the nucleon. By accurately extracting the value of the GLS sum rule from experimental data and comparing it with theoretical predictions, we can gain insights into the quark and parton distributions inside the nucleon and shed light on the physics beyond the Standard Model.

In the work of Ref.~\cite{Leung:1992yx}, authors reported the following result for the
measurement of the GLS sum rule at the scale $|Q^2| = 3$ GeV$^2$,
%-------------------------------------------
\begin{equation} \label{eq:GLSCCFR} 
	\text{GLS} \, ( |Q^2| = 3 \; {\rm GeV}^2) = 2.5 \pm 0.018 \, (\text{stat.}) \pm 0.078 \, (\text{syst.}).
\end{equation}
%-------------------------------------------
The value of the GLS sum rule at the scale $|Q^2|=8$ GeV$^2$ is reported as $2.62 \pm 0.15$ in Ref.~\cite{Londergan:2010cd}. In our work, we obtained $\text{GLS} , (|Q^2|=8 , \text{GeV}^2) = 2.46591\pm0.06289$ for the NLO analysis $\text{GLS} ,(|Q^2|=8 , \text{GeV}^2) =2.46271\pm0.04021$  and for N$^2$LO analysis ,$\text{GLS} ,(|Q^2|=8 , \text{GeV}^2) =2.32743\pm0.03761$ and for N$^3$LO analysis in Mellin space which are in good agreement with the results obtained by mentioned research groups. 
\section{summary and  conclusions}\label{sec:sec8}
In this paper, we present an analysis of the valence quark distribution functions in the proton using Gegenbauer polynomials for their parameterization.  These polynomials have many advantages in QCD analysis; they provide a flexible framework for expanding functions, allowing the approximation of various shapes and behaviors in the PDFs, and facilitating the fitting of experimental data and extraction of relevant physical quantities. Moreover, Gegenbauer polynomials enable the systematic incorporation of higher-order NNNLO corrections, thereby enhancing the precision of theoretical predictions in QCD analyses.  In our analysis, we employ precise splitting functions and Wilson coefficients, eschewing the use of Pad\'e approximation, at the Next-to-Next-to-Next-to-Leading Order. In fact we demonstrated that using the Gegenbauer polynomial for parameterization is a fast, precise, and direct way to calculate the final structure function with high accuracy. The comparison with the MSTH23 and MSTH20 results, which used the Chebyshev polynomials, shows the advantage of the Gegenbauer polynomials and provides a more precise determination of $xF_3$ in the kinematic region of interest without complicated calculations.

 Through careful analysis, the figures presented in our study show a convincing correspondence with both established models and empirical data, attesting to the robustness and reliability of our methodology. This concordance underscores the validity of our findings and supports the wider applicability of our approach within the field of particle physics. 
\appendix

\section{FORTRAN PACKAGE OF OUR
NLO , NNLO and NNNLO PDFS}
A {\tt FORTRAN} package containing our unpolarized PDFs a at NLO , NNLO and NNNLO approximation as
well as the unpolarized structure functions $xF_3(x, Q^2
)$  can be obtained via
Email from the authors upon request. This package includes an example program to illustrate the use of the
routines.

\section*{Acknowledgments}

F.~A. acknowledges the Farhangian University for the provided support to conduct this research. S. A. T.  is grateful to the School of Particles and Accelerators, 
Institute for Research in Fundamental Sciences (IPM).

%\newpage
%\clearpage
%
%%%%%%%%%%%%%%%%%%%%%%%%%%%%%%%%%%%%%%%%%%%%%%%%%%%%%%%%%%%%%%%%%%%%%%%%%%%%%%%%%%%%%%%%%%%%%%%
%

\end{document}